\documentclass[twocolumn,superscriptaddress,prl]{revtex4}
\usepackage[latin9]{inputenc}
\usepackage{amsmath}
\usepackage{amssymb}
\usepackage{verbatim}
\usepackage{graphicx}
\usepackage{color}

\makeatletter

\providecommand{\tabularnewline}{\\}

\@ifundefined{textcolor}{}
{%
 \definecolor{BLACK}{gray}{0}
 \definecolor{WHITE}{gray}{1}
 \definecolor{RED}{rgb}{1,0,0}
 \definecolor{GREEN}{rgb}{0,1,0}
 \definecolor{BLUE}{rgb}{0,0,1}
 \definecolor{CYAN}{cmyk}{1,0,0,0}
 \definecolor{MAGENTA}{cmyk}{0,1,0,0}
 \definecolor{YELLOW}{cmyk}{0,0,1,0}
}

\usepackage{epsfig}\usepackage{bm}\usepackage{mathrsfs}

\renewcommand{\b}[1]{\boldsymbol{#1}}
\newcommand{\gammabar}{\ensuremath\gamma\kern-0.53em-}

\makeatother

\begin{document}

\title{Coherent Transmutation of Electrons into Fractionalized Anyons}

\author{Maissam Barkeshli}

\affiliation{Station Q, Microsoft Research, Santa Barbara, California 93106, USA}

\author{Erez Berg}

\affiliation{Department of Condensed Matter Physics, Weizmann Institute of Science,
Rehovot, 76100, Israel}

\author{Steven Kivelson}

\affiliation{Department of Physics, Stanford University, Stanford, California
94305, USA }
\begin{abstract}
Electrons have three quantized properties -- charge, spin, and Fermi statistics -- that
are directly responsible for a vast array of phenomena. 
Here we show how these properties can be coherently and dynamically
stripped from the electron as it enters certain exotic states of matter
known as a quantum spin liquid (QSL). In a QSL, electron spins collectively
form a highly entangled quantum state that gives rise to emergent
gauge forces and fractionalization of spin, charge, and statistics.
We show that certain QSLs host distinct, topologically robust boundary
types, some of which allow the electron to coherently enter the QSL
as a fractionalized quasiparticle, leaving its spin, charge, or statistics
behind. We use these ideas to propose a number of universal,
conclusive experimental signatures that would establish fractionalization in QSLs. 
\end{abstract}
\maketitle
\textit{Introduction }-- 
A notable example of emergence in physics is \textit{fractionalization}, where the long-wavelength, low-energy
excitations of a quantum phase of matter possess quantum numbers that
are fractions of those of the microscopic constituents. In a fractional
quantum Hall (FQH) state for example, the emergent quasiparticle excitations
carry fractional electric charge and ``anyonic'' quantum statistics.
In a quantum spin liquid (QSL), the electron fractionalizes at low
energies into two quasiparticles -- a spinon and a holon -- that independently
carry the spin and charge of the electron 
\cite{Anderson1, kivelson1987, read1991, wen1991, senthil2000,moessner2001,balents2002,Seidel2009}.
When an electron is injected into such systems, it can decay into
fractionalized components, but a direct quantum mechanical
conversion of an electron to a single fractionalized quasiparticle
has conventionally been thought to be impossible. Consequently, the
question of how to experimentally detect fractionalization in a QSL,
even in principle, has remained a major challenge. 
It is particularly timely to revisit this question, given that there
are now a number of materials that have recently been shown to exhibit anomalous
properties that may indicate that they are QSLs - for a review, see
\cite{balents2010}.

Here we show that some QSLs allow electrons to coherently
enter through their boundary as a fractionalized quasiparticle, leaving
behind their charge, spin, or even Fermi statistics. We
show that this leads to universal experimental signatures that could
provide incontrovertible evidence of fractionalization. 
Our considerations are based on recent theoretical breakthroughs
regarding the physics of two-dimensional (2D) topologically
ordered states associated with extrinsic line and point 
defects \cite{barkeshli2012a,barkeshli2013genon,barkeshli2013defect,beigi2011,kitaev2012,levin2013,kapustin2011,lindner2012,clarke2013,cheng2012},
of which robust boundary phenomena are a special case.

We focus primarily on explaining these phenomena in
the context of the simplest gapped 2D QSL, the $Z_{2}$ short-ranged resonating valence bond state (sRVB)
state, and explaining how it can be distinguished from nonfractionalized
magnetic insulators, or even from QSLs with different types of fractionalization. 
This type of QSL has been proposed\cite{Punk2014} to explain recent neutron scattering
experiments in ZnCu$_3$(OH)$_6$Cl$_2$ (Herbertsmithite)
\cite{Han2012}, and also a number of experimental observations of
the organic compound $\kappa$-(ET)$_2$Cu$_2$(CN)$_3$ \cite{qi2009z2qsl}.
We will build on recent results that
demonstrate that gapped fractionalized phases support topologically
distinct types of gapped boundaries 
\cite{levin2013,barkeshli2013defect}. These
results imply that the $Z_{2}$ sRVB state \textit{necessarily }supports
exactly two topologically distinct types of gapped boundaries, 
referred to as the $e$ edge and the $m$ edge
\footnote{These were identified long ago in the toric code model \cite{kitaev2003}
in terms of rough versus smooth edges \cite{bravyi1998}, although
the generic necessity to realize one or the other in RVB states and
their general independence of lattice structure has only recently
been fully appreciated.
}, that are separated by a topological quantum phase transition.

The $Z_{2}$ sRVB state is an insulating, spin-rotationally 
invariant gapped spin liquid state. That it is a legitimate state of matter has been proven by constructing
model Hamiltonians with exact sRVB ground-states\cite{rokhsarkivelson88,moessner2001,balents2002,senthil2002}.
At low energies, there are four topologically distinct types of elementary
quasiparticle excitations: (i) topologically trivial excitations, which can 
be created with local operators, (ii) spinons and holons, which
carry spin-1/2 and charge 0, or spin 0 and charge 1, respectively,
(iii) visons, which do not carry spin or charge and which have mutually
semionic statistics with respect to the spinons and holons, and (iv)
the composite of a spinon or holon with a vison. Because electrons 
always carry both unit electric charge and spin-1/2, the
spinons and holons cannot individually be created by any local combinations of
electron operators, and therefore must be topological excitations.
That spinons and holons are topologically equivalent follows from the
fact that one can be converted into the other by the local operation
of adding or removing an electron. Moreover, the spinons and holons
can be either bosonic or fermionic, depending on detailed energetics. 

The $Z_{2}$ sRVB state can be understood at low energies in terms
of a $Z_{2}$ lattice gauge theory. In this language, the spinons and
holons carry the $Z_{2}$ gauge charge, while the visons are the $Z_{2}$
fluxes. As a result, the spinons and holons are sometimes referred
to as the $e$ particles and the visons as the $m$ particles. 
This state can also be described at long wavelengths using Abelian
Chern-Simons (CS) field theory \cite{freedman04,wen04}: 
\begin{align}
\mathcal{L}=\frac{1}{4\pi} K^{IJ} \epsilon_{\mu\nu\lambda} a_{\mu}^I \partial_{\nu} a_{\lambda}^J+ \mathcal{L}_{matter},\label{CStheory}
\end{align}
where  $\mu,\nu,\lambda=0,1,2$ are 2+1D space-time indices;
$\epsilon_{\mu\nu\lambda}$ is the Levi-Civita tensor, 
$K = \left(\begin{matrix} 0 & 2 \\ 2 & 0 \end{matrix} \right)$; $I,J = 1,2$;
and $\mathcal{L}_{matter}$ describes the fractionalized quasiparticles, which are minimally coupled to the $U(1)$ gauge
fields $a^I$. The visons carry unit charge under $a^1$, while 
the spinons and holons both carry unit charge under $a^2$. 
The CS term binds charges to fluxes in such a way as to properly capture the nontrivial mutual statistics 
between spinons or holons and visons. 

Clever numerical studies of spin 1/2 frustrated Heisenberg
models\cite{jiang2012a,jiang2012b,yan2011,depenbrock2012}
provided evidence for gapped spin-liquid ground states. 
However, it is not yet clear whether the ground state in those models
is a $Z_2$ sRVB state or a certain competing
QSL, the ``doubled semion'' state, which is characterized instead by
$K =\left(\begin{matrix}2 & 0\\ 0 & -2 \end{matrix}\right)$. 

Each topologically ordered phase, characterized by a matrix $K$, 
can support topologically distinct types of gapped boundaries.
A classification of such gapped edges \cite{levin2013,barkeshli2013defect}, 
when applied to the $Z_2$ sRVB state, predicts exactly two topologically distinct types of gapped edges. As we explain below,
these correspond to whether the $e$ or the $m$ particles are condensed 
along the boundaries. The doubled semion state, in contrast, possesses
only one type of gapped boundary (see Supplemental materials). 

The edge theory can be derived by starting with the Abelian CS theory (\ref{CStheory}).  
[For an alternative explanation, see Supplemental Materials.] 
It is well-known \cite{wen04} that on a manifold with a boundary, Eq. (\ref{CStheory}) is only 
gauge-invariant if the gauge transformations are restricted to be zero on the boundary. This 
implies that on the boundary, the gauge fields correspond directly to physical degrees of 
freedom. One can derive an edge Lagrangian \cite{wen04} in terms of scalar fields $\phi_I$:
\begin{align}
\label{edgeTheory}
\mathcal{L}=\frac{1}{4\pi}K_{IJ}\partial_{x}\phi_{I}\partial_{t}\phi_{J}-V_{IJ}\partial_{x}\phi_{I}\partial_{x}\phi_{J},
\end{align}
where $V_{IJ}$ is a positive-definite velocity matrix. The number of left (or right) movers is given by 
the number of positive (or negative) eigenvalues of the $K$ matrix. In a Hamiltonian formulation, the first
term on the right hand side of Eq. (\ref{edgeTheory}) implies that 
$[\phi_{I}(x),\phi_{J}(y)]=i\pi K_{IJ}^{-1}sgn(x-y)$. Quasiparticles that carry unit charge under 
$a^I$ are created with the operators $e^{i \phi_I}$. 

In the $Z_{2}$ sRVB state, the edge theory maps onto
a single-channel Luttinger liquid, as there are two conjugate fields,
$\phi\equiv\phi_{1}$ and $\theta\equiv\phi_{2}$, with $[\phi(x),\theta(y)]=i\frac{\pi}{2}sgn(x-y)$. 
A composite of two identical quasiparticles in the $Z_2$ sRVB state always corresponds 
to a topologically trivial excitation, and therefore $e^{i 2 \phi_I}$ corresponds to a local operator on the edge.
There are thus two basic types of local terms,
$\delta\mathcal{L}_{Z_{2}}=\lambda_{m}\cos(2\phi)+\lambda_{e}\cos(2\theta)$,
with coupling constants $\lambda_{e,m}$,
that effectively backscatter counterpropagating modes and can induce an 
energy gap on the edge. \footnote{If the resulting gaps are large, the field theory description of
the edge is no longer controlled, but since the most important qualitative
features of a gapped state do not depend on the magnitude of the gap,
the field theory description of such phases should be adequate for
present purposes -- more formally, it gives a leading order description
of the edge modes in the vicinity of an edge quantum critical point
where the edge gaps, but not the bulk gaps are vanishingly small.}

Because $\phi$ and $\theta$ are conjugate, the cosine terms cannot
simultaneously pin their arguments, so there are two distinct phases. 
Where $|\lambda_{m}|$ is the dominant coupling, 
$\langle e^{i\phi}\rangle\neq0$ and $\langle e^{i\theta}\rangle=0$, implying that
the $m$-particles are condensed on the edge. 
Conversely, if $|\lambda_{e}|$ is dominant, the $e$ particles are condensed on the edge:
$\langle e^{i\theta}\rangle\neq0$ and $\langle e^{i\phi}\rangle=0$. 
The two phases, referred to as the $m$ edge and $e$ edge, respectively, 
are topologically distinct. In the absence of any additional global 
symmetries, there is a single quantum critical point between these two gapped 
phases in the Ising universality class. [See the Supplemental Materials for additional discussion.]

In the presence of spin rotation and charge conservation symmetries,
the modes $\phi$ and $\theta$ must represent either low-energy spin
or charge fluctuations (but not both), depending on the physical situation.
If they describe charge fluctuations, then the charge density 
is given by $\rho_{c}=\frac{1}{\pi}\partial_{x}\phi$, and $h=e^{i\theta+in\phi}$
creates a holon which is bosonic for even integer $n$ and fermonic for odd $n$. The
operator $e^{2i\theta}$ is a topologically trivial excitation that
carries charge 2 and no spin and is therefore physically equivalent
to a Cooper pair. Alternatively, if the boson modes describe spin
fluctuations, then the spin density is $S^{z}=\frac{1}{2\pi}\partial_{x}\phi$,
and $S^{\pm}=e^{\pm i2\theta}$. The operator $e^{\pm i\theta+in\phi}$ creates a spin-1/2 
spinon that is bosonic or fermionic for $n$, respectively, even
or odd. In all cases, $e^{i\phi}$ creates a vison. 

\begin{table}
\begin{tabular}{cc}
Edge physics  & Edge type \tabularnewline
\hline 
Spin/charge conserved  & $m$ \tabularnewline
Heisenberg exchange to collinear SDW  & $m$ \tabularnewline
Heisenberg exchange $J>J_{c}$ to N.C. SDW  & $e$ \tabularnewline
Magnetic field $B>B_{c}$ at edge of XXZ system  & $e$ \tabularnewline
Pair tunneling $t_{pair}>t_{pair}^{c}$ to superconductor  & $e$ \tabularnewline
\hline 
\end{tabular}\caption{\label{edgeTable} Summary of conditions under which 
an $e$ or $m$ edge can be realized in the $Z_{2}$ sRVB. Aside from the XXZ case with an
applied magnetic field $B$, the QSL is assumed to be $SO(3)$ spin
rotationally invariant. The first two cases can, 
in principle, also be gapless, instead of realizing the gapped $m$ edge. $J_c$, $B_c$ and $t_{\text{pair}}^c$
are the critical Heisenberg exchange, magnetic field, and pair-tunneling strength needed to realize the 
$e$ edge, respectively. 
}
\end{table}

If charge and spin are conserved, any term proportional to $\cos(2\theta)$ is prohibited,
because this term changes either the charge or the spin of the edge.
It follows that an $e$ edge is incompatible with spin and charge conservation; the $m$-edge is
the generic gapped boundary of a $Z_{2}$ sRVB if charge and spin 
are conserved. 

We will now explore how an $e$ edge can be realized in a physically realistic
system by bringing the edge into contact with another system with one or another pattern of
symmetry breaking. Our results, explained below, are summarized in Table \ref{edgeTable}.
Let us begin by considering a realization of a $Z_{2}$ sRVB state 
with easy-plane, or XXZ, spin-rotational symmetry. 
Again, this can be treated from the perspective of an Ising gauge
theory or from the field theory perspective. 
In the bosonized edge  theory, the bulk $U(1)$ spin rotational
symmetry is associated with the global transformation $\theta\rightarrow\theta+f$ (where $f$
is an arbitrary constant),
and therefore, so long as this symmetry is not explicitly broken, 
terms that would pin $\theta$, such as $\cos(2\theta)$, are disallowed.
However, a magnetic field applied at the edge in an in-plane direction 
leads to a term $-\mu_B B\cos(2\theta)$ (where $\mu_B$ is the Bohr magneton and $B$
is the magnetic field), which for strong enough magnetic
field can produce a phase transition to an $e$ edge.

Now let us consider coupling a paired superconductor to the edge of
a spin liquid. For simplicity we will consider a singlet superconductor,
although the same results apply for the triplet superconductor. At
low energies the dominant couplings between the superconductor and
the spin liquid include Cooper pair tunneling of the form
\begin{align}
\mathcal{H}_{edge}=t_{\text{pair}}(\Phi_{qsl}^{\dagger}\Phi_{sc}+H.c.)
\label{edgeSC}
\end{align}
where $\Phi_{sc}$ is the Cooper pair operator
on the edge of the superconductor, $\Phi_{qsl}$ 
is the Cooper pair operator on the edge of the QSL, $t_{\text{pair}}$ is the pair-tunneling
amplitude, and $H.c.$ stands for Hermitian conjugate. $\Phi_{sc}$
can be replaced by a $c$-number because the Cooper pairs are condensed
in the superconductor. The first term therefore prefers to condense
pairs of holons, which are the Cooper pairs on the spin liquid edge.
In the bosonized field theory, this term corresponds to a perturbation 
$\delta\mathcal{L}_{\text{pair}}\propto t_{\text{pair}}\langle\Phi_{sc}\rangle\cos(2\theta)$,
where the charge density is $\rho_{c}=\frac{1}{\pi}\partial_{x}\phi$.
$t_{\text{pair}}$ can drive a phase transition into the $e$ edge, where single
holons are also condensed: $\langle e^{i\theta}\rangle\neq0$. 
The Cooper pair tunneling does not need to overcome the charge
gap of the $Z_{2}$ spin liquid. By tuning the chemical potential in the superconductor 
using a gate voltage, the energy cost to adding holons to the edge can be made much smaller
than the charge gap. In this situation, even a small pair-tunneling amplitude
is sufficient to condense the holons. 

A similar analysis shows that strong Heisenberg exchange coupling to a non-colinear SDW can also realize
the $e$ edge. Note that coupling an $SO(3)$ spin rotationally
invariant QSL to a colinear SDW (Néel state) is not by itself sufficient to
realize the $e$ edge, because the Néel state has a residual $U(1)$
spin rotation symmetry that precludes spinon condensation.

\begin{figure}
\includegraphics[width=3.2in]{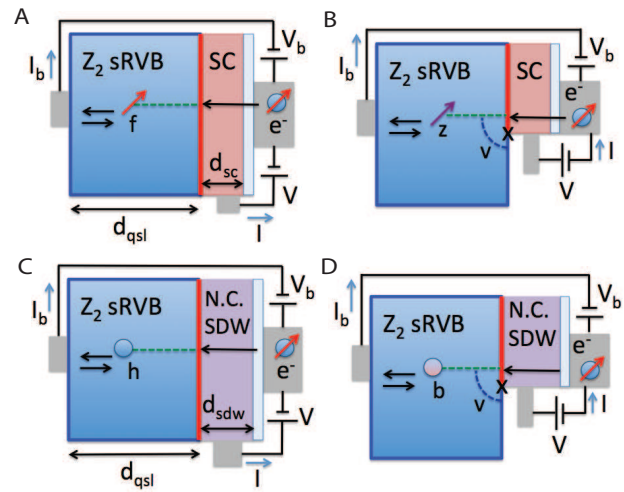} \caption{\label{tomaschFig} 
Proposed geometries to observe geometric resonances
(generalized Tomasch oscillations) in QSLs. $d_{qsl}$ and $d_{sc}$
are the widths of the QSL and superconducting regions, respectively.
(A) An $e$ edge is created by coupling to a superconductor (SC), in which
case an electron can coherently enter the $Z_{2}$ sRVB as a fermionic
spinon $f$. (B) At the domain wall between $e$ and $m$ edges there
is a Majorana fermion zero mode, allowing electrons to coherently
pass into the QSL as a \textit{bosonic }spinon $z$ by emitting a vison
$v$ at the $m$ edge. (C) An $e$ edge is created by coupling a spin-rotationally
invariant QSL to a noncolinear (N.C.) SDW, allowing an electron to coherently
enter the QSL as a fermionic holon, $h$. (D) The Majorana fermion
zero mode at the domain wall allows an electron to coherently enter
the QSL as a bosonic holon, $b$. In (A) to (D), oscillations in $I(V)$
with a period determined by $d_{qsl}$ [without oscillations
in the bulk current $I_{b}(V_{b})$] would provide a conclusive signature
of fractionalization in the QSL.}
\end{figure}

Now that we have investigated the physical conditions under which the distinct edge phases
can be realized, we turn to describing their physical implications.
In the $Z_{2}$ sRVB, the electron $c_{\alpha}$ can be thought of as a composite of a bosonic holon
$b$ and a fermionic spinon $f_{\alpha}$: $c_{\alpha}=bf_{\alpha}$,
where $\alpha=\uparrow,\downarrow$. When the $e$ edge is created
by strong Cooper pair tunneling from a superconductor, the
bosonic holon is condensed on the edge. Consequently, any electron
tunneling term along the boundary between the QSL and the superconductor
becomes
\begin{align}
\delta\mathcal{H}_{edge}=tc_{qsl}^{\dagger}c_{sc}+H.c.=t\langle b_{qsl}\rangle f_{qsl}^{\dagger}c_{sc}+H.c.
\end{align}
Thus, at the $e$ edge with a superconductor, the electron
can coherently tunnel into the spin liquid as a fermionic spinon, leaving its charge
behind at the edge. This would not be allowed at the $m$ edge, at which the electron would 
tunnel into the spin liquid as a whole. Depending on details of the
energetics of the excitations in the QSL, the electron could subsequently
decay into a holon and a spinon.

There is a useful analogy here with Tomasch oscillations observed long ago 
in the context of superconducting films \cite{tomasch1965}. These oscillations
reflect processes in which an electron with energy in excess of the superconducting gap
tunnels coherently from a metal into a superconductor, where it becomes a Bogoliubov
quasi-particle \cite{mcmillan1966}. There is a well-defined sense \cite{kivelsonrokhsar90}
in which the quasi-particles in a conventional superconductor are neutral spinons, although the broken
gauge symmetry of the superconductor makes this analogy somewhat subtle. 

The possibility of a direct coupling between electrons and fractionalized
quasiparticles opens a new realm of possible probes of QSLs. 
Given the existence of a material with a $Z_{2}$ sRVB ground state
and stable fermionic spinon quasi-particles, a direct experimental
signature of such coherent fractionalization of the electron could
be obtained by detecting a suitably generalized version of Tomasch
oscillations: Consider the local electron tunneling density of states (LDOS),
measured at the boundary of the superconductor, for a finite strip
geometry (Fig. \ref{tomaschFig}A). This will receive contributions
from processes where superconducting quasiparticles coherently propagate
into the spin liquid as fermionic spinons, reflect off the outer boundary,
and propagate back. If the spinon inelastic mean free path is larger than the
width $d_{qsl}$ of the QSL, this leads to coherent oscillations of
the LDOS as a function of the dimensionless ratio $eVd_{qsl}/\hbar v_{qsl}$
for voltages $V$ larger than both the spinon and superconducting
gaps, where $v_{qsl}$ is the spinon velocity in the QSL. 
Moreover, incontrovertible evidence that the oscillations are associated
with charge neutral excitations can be obtained by simultaneously
monitoring the bulk current ($I_{b}$ in the figure), 
which in this case will be parametrically small and
free of signatures of coherent spinon interference. Note that for
an $m$ edge, there would be no such coherent oscillations, as the
electron must enter into the QSL as a whole and would subsequently
decay into a spinon and holon.

Similar phenomena will occur when the $e$ edge is created through
magnetic effects, such as through a magnetic field applied at the
edge, or through coupling to a noncolinear SDW. In these cases,
the bosonic spinon is condensed at the edge. We can write the electron
operator as $c_{\alpha}=hz_{\alpha}$, where $z_{\alpha}$ is a bosonic
spinon, and $h$ is a fermionic holon. Now the electron-tunneling
Hamiltonian at the edge becomes $\delta\mathcal{H}_{edge}\propto t_{\alpha\beta}\langle z_{\alpha}\rangle h_{qsl}^{\dagger}c_{\beta;sdw}+H.c.$,
where here we have included a spin-dependent tunneling matrix element
$t_{\alpha\beta}$. Thus, the electron in this case can propagate
coherently into the spin liquid as a fermionic holon. If fermionic
holons are stable fractionalized quasi-particles, we again expect to
observe the Tomasch oscillations in finite strip geometries 
(Fig. \ref{tomaschFig}C).

Treating the case of gapless spin liquids in a theoretically controlled manner is more difficult than the case of gapped spin liquids.
Nevertheless, we expect that in gapless systems such as $Z_{2}$ spin liquids with Dirac points
in their spinon spectrum, or $Z_{2}$ chiral spin liquids with stable spinon Fermi surfaces\cite{Barkeshli2013}, generalized Tomasch oscillations of the
sort envisaged here would occur as well. This is because (i) in such states, the $Z_{2}$ gauge field is gapped, and thus low-energy spinons 
can propagate coherently; and (ii) there is still a well-defined notion of whether the spinons or holons have condensed near the boundary. 

At the boundary between two topologically distinct segments of edge,
localized exotic topological zero modes that give rise to topologically protected degeneracies and projective
non-Abelian statistics \cite{barkeshli2013defect}.
In the case of the $Z_{2}$ sRVB state, the domain wall between 
$e$ and $m$ edges localizes a Majorana fermion zero mode, with the
following physical consequences: Let us consider the case of the superconductivity-induced $e$ edge,
where an electron can coherently enter the QSL as a fermionic spinon.
If this process occurs in the vicinity of the domain wall between
an $e$ and $m$ edge, then the fermionic spinon can also emit or
absorb a vison from the $m$ edge, thus becoming a bosonic spinon.
In other words, the Majorana fermion zero mode is a source or sink of
fermion parity, allowing the electron to coherently enter into the spin liquid as a bosonic spinon.
If the fermionic spinon in the bulk of the QSL can decay into a vison and a bosonic spinon, then this geometry (Fig.
\ref{tomaschFig}B) will allow the Tomasch oscillations to be observed
in the tunneling conductance. Similar considerations show that when
the $e$ edge is induced by magnetism, the electron can enter into
the spin liquid as a bosonic holon in the vicinity of the
$e$-$m$ domain wall (Fig. \ref{tomaschFig}D).

The considerations of this paper suggest ways to tune
through the topological phase transition that separates the $e$ and $m$ edges,
such as by applying a magnetic field to the edge of an easy-plane QSL. 
This can be done by taking a thin sample and shielding the bulk of the QSL by sandwiching
it between two superconductors. At the edge quantum phase transition, there will be enhanced
thermal transport through the edge, leading to a non-zero intercept at low temperatures
in the thermal conductance: $\lim_{T\rightarrow0}\kappa/T=N_{L}c\frac{\pi^{2}}{3}\frac{k_{B}^{2}}{h}$,
where $N_L$ is the number of layers in the QSL, $c = 1/2$ is the central carge of the edge critical point,
and $k_B$ is Boltzmann's constant. Because neither the trivial paramagnet nor the doubled semion QSL 
have topologically distinct types of gapped boundaries, 
the observation of a topological quantum phase transition
at the edge of a gapped insulating spin system 
would prove the existence of a fractionalized spin liquid state, 
and rule out the doubled semion state. The present considerations are readily extended to other sorts of
topologically ordered states, such as those that occur in fractional quantum Hall system. 

\textit{Acknowledgements}-- This work was 
supported by NSF - DMR 1265593 (SK), the Israel Science Foundation (ISF), the Israel-US Binational Foundation (BSF), the Minerva foundation, and the German-Israeli Foundation (GIF) (EB). MB thanks Xiao-Liang Qi for recent collaborations
on related topics, and C. Nayak, M. Freedman, Z. Wang, K. Walker, M. Hastings, 
M.P.A. Fisher, B. Bauer, and T. Grover for discussions. 

\bibliography{TI}

\pagebreak

\appendix

\begin{widetext}
\section{Absence of gapless phase at $Z_2$ sRVB edge}\par In
the main text we argued that generically, in the absence of any global
symmetry, the Luttinger liquid edge theory of the $Z_{2}$ sRVB will
be in a gapped phase. Here we will briefly provide the details of
this analysis. \par Consider the Luttinger liquid edge theory of the
$Z_{2}$ sRVB state, which we write as: 
\begin{align}
\mathcal{L}=\frac{1}{\pi}\partial_{x}\phi\partial_{t}\theta-\frac{u}{2\pi}(K(\partial_{x}\theta)^{2}+\frac{1}{K}(\partial_{x}\phi)^{2}).\label{eq:L}
\end{align}
The two most relevant, allowed perturbations on the edge that lead
to an energy gap are, as explained in the main text, 
\begin{align}
\delta\mathcal{L}=\lambda_{m}\cos(2\theta)+\lambda_{e}\cos(2\phi).\label{eq:dL}
\end{align}
For a Luttinger liquid parameter $K$, the operator $e^{i2\phi}$
has scaling dimension $K$, while $\cos(2\theta)$ has scaling dimension
$1/K$ \cite{giamarchi2003}. 
Therefore, $\lambda_{m}$ is an irrelevant
perturbation when $1/K>2$, while $\lambda_{e}$ is irrelevant when
$K>2$. We see that it is impossible for both to be irrelevant simultaneously.
This implies that unless there is a symmetry which prohibits one of
them to be zero, the gapless Luttinger liquid edge will be generically
unstable to one of the two gapped edge phases.

At $K=1$, the model in Eqs. (\ref{eq:L},\ref{eq:dL}) can be solved
exactly by refermionization 
\cite{shelton1996}. 
One finds a phase transition at $\left|\lambda_{m}\right|=\left|\lambda_{e}\right|$
of the Ising universality class. This is in agreement with the fact
that this model can be viewed as a representation of a two-dimensional
classical planar spin model with a two-fold anisotropy, which is expected
to have similar properties to the two-dimensional Ising model 
\cite{schulz1986}.

Note that the discussion above is not the most general possible analysis,
as we ignore any possible current-density coupling, $\partial_x \theta \partial_x \phi$,
which breaks both inversion symmetry along the edge, $x \rightarrow -x$, 
and time reversal symmetry. We also assume translation symmetry to simplify the
analysis. A more systematic analysis of the possible edge structure will include
disorder as well, and is left for future work. 

\section{$Z$$_{2}$ gauge theory description}

\subsection{$Z_2$ sRVB}

The long-wavelength effective theory of a $Z_{2}$ quantum spin liquid
consists of a $Z_{2}$ gauge theory coupled to spinon and holon matter
fields. Here, we use this effective theory to demonstrate that a transition
from an $m$ edge to an $e$ edge can be induced by breaking either
charge or spin symmetries at the edge.

The following effective Hamiltonian represents a lattice version of
the low-energy theory:

\begin{align}
\mathcal{H} & =-J\sum_{\square}\prod_{\langle
  ij\rangle\in\square}\sigma_{ij}^{z}-\lambda\sum_{\langle
  ij\rangle}\sigma_{ij}^{x}-
\sum_{\langle ij\rangle}\left(t_{b}\sigma_{ij}^{z}b_{i}^{\dagger}b_{j}^{\vphantom{\dagger}}+\sum_{\alpha=\uparrow,\downarrow}t_{s}\sigma_{ij}^{z}s_{i\alpha}^{\dagger}s_{j\alpha}^{\vphantom{\dagger}}\right)\nonumber \\
 & +\sum_{i}\left(m_{b}b_{i}^{\dagger}b_{i}^{\vphantom{\dagger}}+m_{s}\sum_{\alpha=\uparrow,\downarrow}s_{i\alpha}^{\dagger}s_{i\alpha}^{\vphantom{\dagger}}
+ \sum_{\alpha,\beta = \uparrow, \downarrow} \Delta_s \epsilon_{\alpha \beta} s_{i\alpha}^\dagger s_{i \beta}^\dagger + H.c.
\right).\label{eq:Hgauge}
\end{align}

Here, $\sigma_{ij}^{z}$ are Ising degrees of freedom residing on
the links of a two-dimensional square lattice ($\langle ij\rangle$
are nearest neighbor sites), and $b_{i}^{\dagger}$, $s_{i\alpha}^{\dagger}$
are bosonic operators that create a holon or spinon with spin $\alpha=\uparrow,\downarrow$
on site $i$, respectively. $\sum_{\square}$ represents summation
over the plaquettes of the square lattice. $J$ is the energy cost
of a $Z_{2}$ magnetic flux (vison) on a plaquette, and $\lambda$
is the strength of the kinetic term of the gauge field. $t_{b}$,
$t_{s}$ are the hopping amplitudes of the holons and spinons, respectively;
$m_{b}$ and $m_{s}$ are their masses. $\Delta_s$ represents spin singlet pairing of the spinons. 
We will assume that $m_{b,s}$ are sufficiently large, such that the holons and spinons are gapped
in the bulk. In addition to $U(1)$ spin and charge conservation,
the Hamiltonian (\ref{eq:Hgauge}) is invariant under local gauge
transformations: $[\mathcal{H},U_{i}]=0$, where 
$U_{i}=\left(\prod_{j\in+_{i}}\sigma_{ij}^{x}\right)e^{i\pi(b_{i}^{\dagger}b_{i}+\sum_{\alpha}s_{i\alpha}^{\dagger}s_{j\alpha})}$
($j$ is a nearest neighbor site of $i$). We work in the sector $U_{i}=1$;
in this sector, a $Z_{2}$ version of Gauss' law is obeyed on every
site.

Let us analyze the model (\ref{eq:Hgauge}) in the limit $\lambda\ll J$,
i.e., deep in the $Z_{2}$ QSL phase. In this limit, to zeroth order
in $\lambda/J$, there are no magnetic fluxes in the ground state,
i.e. $\prod_{\langle ij\rangle\in\square}\sigma_{ij}^{z}=1$ for every
plaquette. Consider a cylindrical system with $L_{x}\times L_{y}$
sites and periodic boundary conditions along the $y$ direction. As
discussed in the main text, the gapped edges of the cylinder at $x=0,L_{x}$
are of the $m-$type. There are two degenerate ground states, corresponding
to sectors with an even or odd magnetic flux through the holes of
the cylinder. In the sector where the magnetic flux through the hole
of the cylinder is zero, the we can always find a gauge for which
$\sigma_{ij}^{z}=+1$ for all $\langle i,j\rangle$. (In the other
sector with one magnetic flux through the hole, one can find a gauge
in which all the $\sigma_{ij}^{z}=+1$, except along a single horizontal
row.)

We now consider spin and charge conservation breaking perturbations
on the edge. A proximity coupling of a superconductor to the edge
can be represented by the term $\Delta\mathcal{H}_{\mathrm{SC}}=-\Delta\sum_{j\in\mathrm{edge}}\left[\left(b_{j}^{\dagger}\right)^{2}+\mathrm{H.c.}\right]$,
creating a pair of charge $e$ holons at the edge. The low-energy
Hamiltonian describing the edge is then

\begin{equation}
\mathcal{H}_{\mathrm{edge}}=\sum_{j}\left[m_{b}b_{j}^{\dagger}b_{j}^{\vphantom{\dagger}}-t_{b}b_{j}^{\dagger}b_{j+\hat{y}}^{\vphantom{\dagger}}-\Delta\left(b_{j}^{\dagger}\right)^{2}+H.c.\right],\label{eq:Hedge}
\end{equation}
where the summation runs over sites along the edge. Note that, after
fixing the gauge, the local gauge invariance of the original Hamiltonian
(\ref{eq:Hgauge}) becomes a global gauge invariance under $b_{j}\rightarrow-b_{j}$
for all $j$. The Hamiltonian (\ref{eq:Hedge}) can be diagonalized
by a Bogoliubov transformation. Upon increasing $\Delta$, we find
that the gap on the edge closes at $2\Delta=m_{h}-2\left|t_{b}\right|$.
The vanishing of the gap signals a condensation of the holon field
$b_{j}$. At larger values of $\Delta$, the spectrum of (\ref{eq:Hedge})
contains imaginary frequencies, and interactions between the holons
{[}neglected in Eq. (\ref{eq:Hgauge}){]} have to be included to retrieve
stability. Presumably, the large $\Delta$ phase is a broken symmetry
phase, in which $\langle b_{j}^{\dagger}b_{j+l}\rangle\rightarrow\mathrm{const.}$
in the limit $l\rightarrow\infty$.$ $ This is the $e-$edge with
condensed bosonic holons.

An analogous treatment in the case of an applied Zeeman field perpendicular
to the $z$ axis shows that for a sufficiently strong field, a transition
to an $e-$edge with condensed bosonic spinons occurs. 

\subsection{Doubled Semion}
\label{dsIsing}

It has recently been pointed out that the doubled semion state, described in the main text in terms of 
the matrix $K = \left( \begin{matrix} 2 & 0 \\ 0 & -2 \end{matrix} \right)$, can be understood in terms
of $Z_2$ lattice gauge theory 
\cite{levin2012}. 
To understand this description, consider an interacting bosonic system 
which forms a $Z_2$ symmetry-protected topological (SPT) state 
\cite{chen2012spt}. 
This is a topologically non-trivial gapped Ising paramagnet, which is distinct from the conventional Ising paramagnet, and which has the property
that as long as the Ising $Z_2$ symmetry is preserved, the edge has robust counterpropagating gapless modes. 
In other words, in the $Z_2$ SPT state, the only way to have a gapped edge is for the $Z_2$ global symmetry
to be broken on the edge. 

The doubled semion state can be understood as a state where the global $Z_2$ symmetry of the $Z_2$ SPT state
has been promoted to a local, gauge symmetry. The gapped edge of the doubled
semion state must therefore necessarily break the $Z_2$ gauge symmetry. That is, in contrast to the $Z_2$ sRVB state
discussed above, the doubled semion state cannot support a gapped edge that preserves the $Z_2$ gauge symmetry. 
Therefore the doubled semion state has only one topological class of
gapped edge. 

\section{Gapless Spin Liquids}

Here we provide a more extended discussion of the case where the quantum spin liquid of interest 
is gapless. We focus on the case where the low energy effective theory of the spin liquid is described by a
$Z_2$ gauge field, coupled to gapless spinons. If the spinons are fermionic, they could form Dirac nodes, as in a $d$-wave 
superconductor, or they could form a Fermi surface. The latter case is stable as long as time-reversal
and lattice inversion symmetry is broken. Such states have been proposed to be realized in several different materials. 

In the language of the previous section, the states with fermionic spinons can be described in terms of a lattice version of the low energy effective theory
as
\begin{align}
\mathcal{H}  =&-J\sum_{\square}\prod_{\langle
  ij\rangle\in\square}\sigma_{ij}^{z}-\lambda\sum_{\langle
  ij\rangle}\sigma_{ij}^{x}-
\sum_{ij}\left(t^b_{ij} \sigma_{ij}^{z}b_{i}^{\dagger}b_{j}^{\vphantom{\dagger}}+
\sum_{\alpha=\uparrow,\downarrow}t^f_{ij} \sigma_{ij}^{z}
f_{i\alpha}^{\dagger}f_{j\alpha}^{\vphantom{\dagger}} +
\sum_{\alpha,\beta = \uparrow, \downarrow} \Delta_{ij;\alpha\beta} f_{i\alpha}^\dagger f_{j\beta}^\dagger + H.c. \right)\nonumber \\
 & +\sum_{i}\left(m_{b}b_{i}^{\dagger}b_{i}^{\vphantom{\dagger}} - \mu_f f_i^\dagger f_i \right)
\end{align}
As before, $b$ represents the bosonic holes, and here we have considered the case where the stable 
spinons are fermionic, described by $f_\alpha$. Depending on the band structure and chemical potential $\mu_f$ 
of the fermions, they may form either a Fermi surface or a set of
Dirac cones. Since the gauge field is $Z_2$, pairing terms $\Delta_{ij;\alpha \beta}$ are allowed in the effective theory. 
Spin rotational symmetry requires the pairing terms to be spin singlet. 

Let us consider coupling the edge of such a system to a superconductor, by adding a term
\begin{align}
\delta \mathcal{H}_{edge} = t_{pair} \Phi_{sc} \sum_{i\in \text{edge}} b_i^\dagger b_i^\dagger + H.c. - \mu \sum_{i\in \text{edge}} b_i^\dagger b_i,
\end{align}
where $i$ sums over the sites near the edge. $t_{pair}$ is the amplitude for tunneling Cooper pairs onto the edge, and $\Phi_{sc}$ is the pair amplitude in the superconductor. $\mu$ is a chemical potential for the edge, which can be tuned by changing the chemical potential in the superconductor. 

It is clear that in such a theory, the bosonic holons can either be condensed ($\langle b \rangle \neq 0$), or uncondensed 
($\langle b \rangle = 0$) on the edge. In the former case, the $Z_2$ gauge symmetry is broken near the edge, while in the latter case it is preserved. These two cases cannot be distinguished by any local order parameter. The effective dynamics on the edge can be analyzed in detail by integrating out the bulk fields, leading to an effective one-dimensional theory for the edge. Since the bulk is gapless, the effective 1D edge theory would be dissipative and contain long-range interactions. Nevertheless, the possibility of two phases, depending on whether the $Z_2$ gauge symmetry is broken or not, is still sharply defined. If the holons are condensed on the edge, then from the discussion in the main text it is clear that electrons can coherently tunnel into the spin liquid as a fermionic spinon. Since the gauge field is gapped, the fermionic spinons can propagate coherently as quasiparticles in the bulk. This should then allow the possibility of the generalized Tomasch oscillations discussed in the main text. 

It is straightforward to adapt the above discussion to the case where bosonic spinons can condense on the boundary. In the above theory, the bosonic spinon corresponds to the composite of a vison and a fermionic spinon. Depending on whether it is condensed on the edge, the $Z_2$ gauge symmetry could be broken or preserved on the edge. 

\section{General Topological Classification of Gapped Edges}

Here we will briefly review some recent general results regarding the classification of general gapped edges in topological phases of matter. A general Abelian topological phase of matter can be described by Chern-Simons gauge theory, characterized by a matrix $K$, as shown in the main text. In this theory, topologically non-trivial quasiparticles are characterized by integer vectors, $\b{l}$, such that two quasiparticles $\b{l}$, $\b{l}'$ are topologically equivalent if $\b{l} = \b{l}' + K \b{\Lambda}$, where $\b{\Lambda}$ is also an integer vector. The mutual statistics between two quasiparticles labelled by $\b{l}$ and $\b{l}'$ is $\theta_{\b{l} \b{l}'} = 2\pi \b{l}^T K^{-1} \b{l}'$. 

Remarkably, Abelian topological phases can have topologically distinct
types of gapped boundaries 
 \cite{kitaev2012,levin2013,barkeshli2013genon,barkeshli2013defect} (see also
 \cite{kapustin2011,fuchs2012}). 
It has been shown recently that these are in one-to-one correspondence with certain subgroups of fractionalized quasiparticles, called Lagrangian subgroups. A Lagrangian subgroup $M$ is a maximal subgroup of mutually bosonic quasiparticles that satisfies two conditions: (1) For every $\b{m},~\b{m}'\in M$, $e^{i\theta_{\b{m}\b{m}'}}=1$, and (2) For every $\b{l}\notin M$, there exists some $\b{m}\in M$ such that $e^{i\theta_{\b{l}\b{m}}}\neq1$. The first condition guarantees that all particles in $M$ are mutually local with respect to each other, while (2) guarantees that every particle not in $M$ has non-trivial statistics with respect to at least one particle in $M$. The physical characteristic of a gapped boundary characterized by $M$ is that the topological quasiparticles in $M$ can be created or annihilated near the boundary by \it local \rm operators, even though they
are non-local topological excitations in the bulk of the system 
\cite{levin2013,barkeshli2013defect}.

As an example, the $Z_{2}$ sRVB state has two different Lagrangian subgroups, $\left\{\left( \begin{matrix} 0 \\ 0\end{matrix} \right) , \left( \begin{matrix} 1 \\ 0\end{matrix} \right) \right\}$
and $\left\{\left( \begin{matrix} 0 \\ 0\end{matrix} \right) , \left( \begin{matrix} 0 \\ 1\end{matrix} \right) \right\}$, and therefore has two topologically distinct types of
gapped boundaries. These are the $e$ and $m$ edges described in the main text. The doubled semion theory has only one Lagrangian subgroup, corresponding to the quasiparticles
$\left\{\left( \begin{matrix} 0 \\ 0\end{matrix} \right) , \left( \begin{matrix} 1 \\ 1\end{matrix} \right) \right\}$, and therefore has only one type of gapped boundary, as explained through the language of Ising gauge theory in Sec. \ref{dsIsing} above. 
\end{widetext}

\end{document}